\documentstyle[12pt]{article}
\catcode`@=11
\newcount\@tempcntc
\def\@citex[#1]#2{\if@filesw\immediate\write\@auxout{\string\citation{#2}}\fi
  \@tempcnta\z@\@tempcntb\m@ne\def\@citea{}\@cite{\@for\@citeb:=#2\do
    {\@ifundefined
       {b@\@citeb}{\@citeo\@tempcntb\m@ne\@citea\def\@citea{,}{\bf ?}\@warning
       {Citation `\@citeb' on page \thepage \space undefined}}%
    {\setbox\z@\hbox{\global\@tempcntc0\csname b@\@citeb\endcsname\relax}%
     \ifnum\@tempcntc=\z@ \@citeo\@tempcntb\m@ne
       \@citea\def\@citea{,}\hbox{\csname b@\@citeb\endcsname}%
     \else
      \advance\@tempcntb\@ne
      \ifnum\@tempcntb=\@tempcntc
      \else\advance\@tempcntb\m@ne\@citeo
      \@tempcnta\@tempcntc\@tempcntb\@tempcntc\fi\fi}}\@citeo}{#1}}
\def\@citeo{\ifnum\@tempcnta>\@tempcntb\else\@citea\def\@citea{,}%
  \ifnum\@tempcnta=\@tempcntb\the\@tempcnta\else
   {\advance\@tempcnta\@ne\ifnum\@tempcnta=\@tempcntb \else \def\@citea{--}\fi
    \advance\@tempcnta\m@ne\the\@tempcnta\@citea\the\@tempcntb}\fi\fi}
\catcode`@=12
\newcommand\mini{j_{\rm min}}
\newcommand\one{1\kern-2.5pt{\rm l}}
\newcommand\Frac[2]{\hbox{$\frac{#1}{#2}$}}
\newcommand{\slv}{v\kern-5pt\raise1pt\hbox{$\scriptstyle/$}\kern1pt}
\begin{document}
\begin{flushright}
MZ-TH/96-09 \\[-0.2cm]
March 1996 \\[-0.2cm]
\end{flushright}
\begin{center}

{\Large\bf Issues in Heavy Flavour Baryons} \\[1.75cm]

{\large J\"urgen G.~K\"orner\footnote{Supported in part by BMBF,FRG
under contract 06MZ566}} \\[.4cm]
Institut f\"ur Physik, Johannes Gutenberg-Universit\"at \\
Staudingerweg 7, D-55099 Mainz, Germany.

\end{center}
\vspace{1.5cm}
\centerline {\bf ABSTRACT}\noindent
I present a mini-review on the physics of heavy flavour baryons where
I concentrate on the HQET description of their exclusive decay modes.
In particular
I discuss the structure of current-induced bottom baryon to charm baryon 
transitions, and the structure of pion and photon transitions between heavy 
charm or bottom baryons in the Heavy Quark Symmetry limit as
${m_Q\rightarrow\infty}$. The emphasis is on the structural similarity of 
the Heavy Quark Symmetry predictions for the three types of transitions. 
The requisite coupling expressions are discussed both in the
covariant framework as well as in terms of
Clebsch-Gordan coefficients and 6-$j$ symbols. At the end of my review
I touch on some unresolved issues in exclusive nonleptonic charm and
bottom baryon decays which serve to highlight our present lack of
understanding of nonleptonic heavy baryon decays.

\newpage

\section{Introductory Remarks}

Because of the initial abundance of data on heavy charm and bottom mesons 
the attention of experimentalists and theoreticians had initially been 
drawn towards applications of the Heavy Quark Effective Theory (HQET) 
to the meson sector. In the meantime the situation has considerably changed 
and data on heavy baryons and their decay properties are starting to become 
available in impressive amounts.
In the charm sector the states $\Lambda_c(2285)$ and
$\Sigma_c(2453)$ are well established while there is first evidence for
the $\Sigma_c^*(2510)$ state. Two excited states $\Lambda_c^{**}(2593)$ 
and $\Lambda_c^{**}(2627)$ have been seen which very likely correspond to 
the two lowest lying $p$-wave excitations of the light diquark system
making up the $\Lambda_{cK1}^{**}$ Heavy Quark
Symmetry (HQS) doublet to be discussed in Sec.2.
The charm-strangeness states $\Xi_c(2470)$ and $\Omega_c(2720)$ as well as 
the $\Xi_c^*(2643)$ have been seen. First evidence was presented for the 
$J^P=\frac{1}{2}^+$ state $\Xi'_c(2570)$  with the flavour configuration 
$c\{sq\}$. Thus almost all ground state charm baryons have been seen 
including two $p$-wave states. In the bottom sector the $\Lambda_b(5640)$
has been identified as well as the $\Sigma_b(5713)$ and the 
$\Sigma_b^*(5869)$. Some indirect evidence has been presented for the 
$\Xi_b(5800)$.

Apart from the fact that the existence of the above heavy baryon states has
now been established there are also copious experimental data on the
production characteristics of heavy baryons and on their exclusive
and inclusive decays.
Most of the data accumulated so far are on charm baryons.
A multitude of different experiments both at collider and fixed target
facilities have contributed to our present knowledge of charm baryon
physics. New experiments are being planned or have already been set up.
For example the SELEX experiment E781 at Fermilab is waiting for
beam time and plans to log $5\times 10^4$ fully reconstructed 
$\Lambda_c \rightarrow p K^+ \pi^-$ decays per year with many
other decay modes reconstructed.
Most of the planned experiments will also see bottom baryons albeit at
somewhat reduced rates. The next decade will see the opening of a number
of new facilities and experiments among which are HERA-B, LHC-B, COMPASS,
CLEO III, CHARM 2000, the US and Japanese B-Factories , and the
$\tau-$charm factory project. Heavy baryon physics may not be the prime
objective of all of these projects but heavy baryons will certainly be seen at
these facilities if only as welcome by-products.   

Let us try and gain a historical perspective on how heavy baryon
production (and detection) rates
have developed over the past years by taking a look at the strangeness sector.
In 1964 V.E.Barnes et al. reported on the first observation
of a single $\Omega^-$ in the BNL 80-in. bubble chamber. Compare this to
the $4\times 10^5$ $\Omega^-$ events recently recorded by the E800 
Collaboration at Fermilab. 
Another impressive rate figure is the projected total of $10^9$ reconstructed
$\Xi$ hyperons at the planned HYPERCP experiment E-871 at Fermilab. If
these figures can be taken as a foreboding of what lies ahead of us
in the charm and bottom baryon sector we are certainly heading for
exciting times. As a theoretician I must ruefully admit, though, that our
understanding of the dynamics of heavy baryons is far from complete. 
With all the heavy baryon data expected to come up in the next future
there is the acute danger that the experimentalists
get ahead of us theoreticians.

The framework to treat the dynamics of heavy baryons is
Heavy Quark Effective Theory (HQET). In Secs. 2, 3 and 4 we develop in some    
detail the leading order HQET
description of semileptonic, one-pion and photon decays of heavy baryons.
The emphasis is on the structural similarity of 
the HQS description of these decays. In Sec. 5\quad I discuss possibilities
to further constrain the Heavy Quark 
Symmetry (HQS) structure of the three type of decays by resolving the light 
diquark transitions in terms of a constituent quark model description of 
the light diquark transitions with an underlying $SU(2N_f)\otimes O(3)$ 
symmetry. Secs. 6 and 7 contain a brief discussion of some multifarious
aspects of exclusive nonleptonic heavy baryon decays.

\section{Heavy Baryon Spin Wave Functions}

Let us begin by constructing the heavy baryon spin wave functions that
enter into the description of heavy baryon decays. A heavy baryon is made 
up of a light diquark system $(qq)$ and a heavy quark $Q$. The light 
diquark system has bosonic quantum numbers $j^P$ with total angular 
momentum $j=0,1,2 \dots$ and parity $P=\pm 1$. To each diquark system with 
spin-parity $j^P$ there is a degenerate heavy baryon doublet with 
$J^P=(j\pm\Frac12)^P$ ($j=0$ is an exception). It is important to realize 
that the HQS structure of the heavy baryon states is entirely determined 
by the spin-parity $j^P$ of the light diquark system. The requisite angular 
momentum coupling factors can be read off from the coupling scheme
\begin{equation}\label{eqn1}
j^P\otimes\Frac12^+\Rightarrow J^P.
\end{equation}
Apart from the angular momentum coupling factors the dynamics
of the light system is completely decoupled from the heavy quark. 

Let us cast these statements into a covariant framework in which the heavy
baryon wave function $\Psi$ describes the amplitude of finding the light 
diquark system and the heavy quark in the heavy baryon. The covariant 
equivalent of the coupling scheme Eq.~(\ref{eqn1}) is then given by
\begin{equation}\label{eqn2}
\Psi=\phi_{\mu_1\cdots\mu_j}\psi^{\mu_1\cdots\mu_j},
\end{equation}
where $\phi_{\mu_1\cdots\mu_j}$ stands for the tensor representation of 
the spin-parity $j^P$ diquark state and $\psi^{\mu_1\cdots\mu_j}$ represents 
the heavy-side baryon spin wave function (in short: heavy baryon wave 
function) coupling the heavy quark to the heavy baryon. Let us be more 
specific. If
\begin{equation}\label{eqn3}
|J^P,m_J\rangle=\sum_{m_j+m_Q=m_J}\!\!\!\!\!\!
  \langle j^P,m_j;\Frac12^+,m_Q|J^P,m_J\rangle
  |j^P,m_j\rangle|\Frac12^+,m_Q\rangle
\end{equation}
defines the light diquark-heavy quark rest-frame wave function,
the C.G. coefficients
determining the heavy quark - light diquark content of the heavy
baryon can be obtained in covariant fashion from the heavy baryon
spin wave function by the covariant projection
\begin{equation}\label{eqn4}
\langle j^P,m_j;\Frac12^+,m_Q|J^P,m_J\rangle
  =\varepsilon^*_{\mu_1\cdots\mu_j}(m_j)\bar u(m_Q)
  \psi^{\mu_1\cdots\mu_j}(m_J).
\end{equation}
The r.h.s. of Eq.~(\ref{eqn4}) can be evaluated for any velocity
four-vector $v_\mu$ of 
the heavy baryon which, at leading order, equals the velocity of the heavy 
quark and the diquark system. Details including questions of normalization 
can be found in~\cite{santafe1}. Differing from~\cite{santafe1} I have 
normalized the spinors appearing in Eq.~(\ref{eqn4}) to $1$ and not to $2M$ 
and $2M_Q$ as in~\cite{santafe1}. It is not difficult to construct the 
appropiate heavy baryon spin wave functions using the heavy quark on-shell 
constraint $\slv\psi^{\mu_1\cdots\mu_j}=\psi^{\mu_1\cdots\mu_j}$ and the 
appropiate normalization condition. In Table 1 (fourth column) I have 
listed a set of correctly normalized heavy baryon spin wave functions that 
are associated with the diquark states $j^P=0^+,1^+,0^-,1^-,2^-$.

Next I turn my attention to the question of which low-lying heavy
baryon states can be expected to exist. From our experience with light 
baryons and light mesons we know that one can get a reasonable description 
of the light particle spectrum in the constituent quark model picture. This 
is particularly true for the enumeration of states, their spins and their 
parities. As much as we know up to now, gluon degrees of freedom do not 
seem to contribute to the particle spectrum. It is thus quite natural to 
try the same constituent approach to enumerate the light diquark states, 
their spins and their parities. 

From the spin degrees of freedom of the two light quarks one obtains a 
spin~0 and a spin~1 state. The total orbital state of the diquark system 
is characterized by two angular degrees of freedom which I take to be the 
two independent relative momenta $k=\frac12(p_1-p_2)$ and 
$K=\frac12(p_1+p_2-2p_3)$ that can be formed from the two light 
quark momenta $p_1$ and $p_2$ and the heavy quark momentum $p_3$. 
The $k$-orbital momentum describes relative orbital excitations of the two 
quarks, and the $K$-orbital momentum describes orbital excitations of the 
center of mass of the two light quarks relative to the heavy quark. The 
$(k,K)$-basis is quite convenient in as much as it allows one to classify 
the diquark states in terms of $SU(2N_f)\otimes O(3)$ representations as 
will be discussed later on. Table~1 lists all ground state $s$-wave and 
excited $p$-wave heavy baryon wave functions as they occur in the
constituent approach to the 
light diquark excitations. They are grouped together in terms of 
$SU(2N_f)\otimes O(3)$ representations with $N_f=2$ for $(u,d)$. The $s$-wave 
states are in the $\underline{10}\otimes\underline{1}$ representation, and 
the $p$-wave states are in the $\underline{10}\otimes\underline{3}$ and 
$\underline{6}\otimes\underline{3}$ representation of $SU(4)\otimes O(3)$ 
for the $K$- and $k$-multiplets, respectively. Apart from the ground state 
$s$-wave baryons one thus has altogether seven $\Lambda$-type $p$-wave 
states and seven $\Sigma$-type $p$-wave states. The analysis can easily be 
extended to the case $SU(6)\otimes 0(3)$ bringing in the strangeness quark 
in addition.

\section{Generic Picture of Current, Pion and Photon Transitions}

In Fig.~1 we have drawn the generic diagrams that describe $b\rightarrow c$ 
current transitions, and $c\rightarrow c$ pion and photon transitions 
between heavy baryons in the HQS limit. The heavy-side and light-side 
transitions occur completely independent of each other (they ``factorize'') 
except for the requirement that the heavy side and the light side have the 
same velocity in the initial and final state, respectively, which are also 
the velocities of the initial and final heavy baryons. The $b\rightarrow c$ 
current transition induced by the flavour-spinor matrix~$\Gamma$ is hard 
and accordingly there is a change of velocities $v_1\rightarrow v_2$, 
whereas there is no velocity change in the pion and photon transitions. 
The heavy-side transitions are completely specified whereas the light-side 
transitions $j_1^{P_1}\rightarrow j_2^{P_2}$,
$j_1^{P_1}\rightarrow j_2^{P_2}+\pi$ and
$j_1^{P_1}\rightarrow j_2^{P_2}+\gamma$ are described by a number of form
factors or coupling factors which parametrize the light-side transitions.
The pion and the photon couple only to the light side. In the case of the
pion this is due to its flavour content. In the case of the photon the
coupling of the photon to the heavy side involves a spin flip which is down
by $1/m_Q$ and thus the photon couples only to the light side in the Heavy
Quark Symmetry limit.

Referring to Fig.~1\quad I am now in the position to write down the generic
expressions for the current, pion and photon transitions according to the
spin-flavour flow depicted in Fig.~1. One has ($\omega=v_1\cdot v_2$)
\medskip\\
{\it current transitions:}
\begin{equation}\label{eqn5}
\bar\psi_2^{\nu_1\cdots\nu_{j_2}}\Gamma\psi_1^{\mu_1\cdots\mu_{j_1}}
\left(\sum_{i=1}^Nf_i(\omega)
t^i_{\nu_1\cdots\nu_{j_2};\mu_1\cdots\mu_{j_1}}\right)
\end{equation}
\begin{eqnarray}
n_1\cdot n_2&=&1\qquad N=\mini+1\nonumber\\
n_1\cdot n_2&=&-1\quad N=\mini\nonumber
\end{eqnarray}
{\it pion transitions:}
\begin{equation}\label{eqn6}
\bar\psi_2^{\nu_1\cdots\nu_{j_2}}\psi_1^{\mu_1\cdots\mu_{j_1}}
\left(\sum_{i=1}^Nf_i^\pi
t^i_{\nu_1\cdots\nu_{j_2};\mu_1\cdots\mu_{j_1}}\right)
\end{equation}
\begin{eqnarray}
n_1\cdot n_2&=&1\qquad N=\mini\nonumber\\
n_1\cdot n_2&=&-1\quad N=\mini+1\nonumber
\end{eqnarray}
{\it photon transitions:}
\begin{equation}\label{eqn7}
\bar\psi_2^{\nu_1\cdots\nu_{j_2}}\psi_1^{\mu_1\cdots\mu_{j_1}}
\left(\sum_{i=1}^Nf_i^\gamma
t^i_{\nu_1\cdots\nu_{j_2};\mu_1\cdots\mu_{j_1}}\right)
\end{equation}
\begin{eqnarray}
j_1&=&j_2\qquad N=2j_1\nonumber\\
j_1&\neq&j_2\quad N=2\mini+1\nonumber
\end{eqnarray}
where the $\psi^{\mu_1\cdots\mu_j}$ are the heavy baryon spin wave
functions introduced in Sec.~2. The pattern of the above decomposition 
parallels the corresponding decomposition in lepton-hadron interactions
where the transition amplitude is written as
$j^\mu_{lepton}\cdot J^{hadron}_\mu$. The structure of the leptonic
current $j^\mu_{lepton}$ is known and the unknown hadronic current
$J^{hadron}_\mu$ is expanded along a set of covariants with the familiar
invariant amplitudes as coefficient functions.

In each of the above cases we have also given the result of counting the
number~$N$ of independent form factors or coupling factors. These are easy
to count by using either helicity amplitude counting or $LS$ partial wave
amplitude counting. In the case of current and pion transitions the
counting involves the normalities of the light-side diquarks which is
defined by $n=(-1)^jP$.

The tensors 
$t^i_{\nu_1\cdots\nu_{j_2};\mu_1\cdots\mu_{j_1}}$ appearing in 
Eq.~(\ref{eqn5}) have to be build from the vectors $v_1^{\nu_i}$ and 
$v_2^{\mu_i}$, the metric tensor $g_{\mu_i\mu_k}$, the pion or the photon
momentum and, depending on parity, 
from Levi-Civita objects such as $\varepsilon(\mu_i\nu_kv_1v_2)
:=\varepsilon_{\mu_i\nu_k\alpha\beta}v_1^\alpha v_2^\beta$. The number 
of independent tensors that can be written down in each of the three cases 
is necessarily identical to the numbers listed in Eqs.~(\ref{eqn5}), 
(\ref{eqn6}) and~(\ref{eqn7}). Lack of space prevents us from giving the 
explicit forms of these tensors. They can be found in~\cite{santafe1}.

The generic expressions Eq.~(\ref{eqn5}), Eq.~(\ref{eqn6}) and 
Eq.~(\ref{eqn7}) completely determine the HQS structure of the current, 
pion and photon transition amplitudes. It is not difficult to work out 
relations between rates, angular decay distributions etc. from these 
expressions.

\section{$6-j$ Symbols in Heavy Baryon Transitions}
It is well worth mentioning that all three covariant coupling expressions
in Sec.3 (current,pion,photon)
can also be written down in terms of Wigner's 6-$j$ symbol 
calculus~\cite{santafe1,santafe2} as can be appreciated from the discussion 
in Sec.~2 (see Eqs.~(\ref{eqn2}) and~(\ref{eqn3})). For example, looking 
at the pion transition in Fig.~1 one sees that one has to perform 
altogether three angular couplings. They are
\begin{eqnarray}
\mbox{(i)}&&\hspace{1.7cm}{j_1}^{P_1}\otimes\Frac12^+\Rightarrow {J_1}^{P_1}
  \nonumber\\
\mbox{(ii)}&&\hspace{1.7cm}{j_2}^{P_2}\otimes\Frac12^+\Rightarrow {J_2}^{P_2}\\
\mbox{(iii)}&&\hspace{1.7cm}{J_2}^{P_2}\otimes L_\pi\Rightarrow {J_1}^{P_1}
  \nonumber
\end{eqnarray}
where $L_\pi=l_\pi$ is the orbital momentum of the pion and ${J_1}^{P_1}$
and ${J_2}^{P_2}$ denote the $J^P$ quantum numbers of the initial and final
baryons. This is a coupling problem well-known from atomic and nuclear 
physics and the problem is solved by Wigner's 6-$j$ symbol calculus. One 
finds~\cite{santafe1,santafe2}
\begin{eqnarray}
\lefteqn{M^{\pi}(J_1J_1^z\rightarrow J_2J_2^z+L_{\pi}m)}\label{eqn8}\\
  &=&M_{L_\pi}(-1)^{L_\pi+j_2+\Frac12+J}(2j_1+1)^{1/2}(2J_2+1)^{1/2}
  \nonumber\\&&
  \qquad\left\{\begin{array}{ccc}j_2&j_1&L_\pi\\J&J_2&\Frac12\end{array} 
  \right\}\langle LmJ_2J_2^z|J_1J_1^z\rangle,\nonumber
\end{eqnarray}
where the expression in curly brackets is Wigner's 6-$j$ symbol and 
$\langle L_\pi MJ_2J_2^z|J_1J_1^z \rangle$ is the Clebsch-Gordan coefficient 
coupling $L_\pi$ and $J_2$ to $J_1$. $M_{L_\pi}$ is the reduced amplitude of 
the one-pion transition. It is proportional to the invariant coupling 
$f_{l_\pi}$ occurring in the covariant expansion in Eq.~(\ref{eqn6}). 

%\begin{figure}
%\bec 
%\input{dubdub.pstex_t}
%\vspace{0.4mm}
%\fcaption{One-pion transition strengths for the transitions $\{
%  \Lambda_{QK2}^{**}\} \rightarrow \{ \Sigma_Q\} + \pi$. Degeneracy levels are
%  split for illustrative purposes.}
%\eec
%\end{figure}

Let us, for example, calculate the doublet to doublet transition rates for
e.g. $\{\Lambda_{Qk2}^{**}\}\rightarrow\{\Sigma_Q\}+\pi$. The rates are in
the ratios $4:14:9:9$ as represented in Fig.~2~\cite{santafe1,santafe3}.
This result can readily be calculated using the 6-$j$ formula 
Eq.~(\ref{eqn8}) and some standard orthogonality relations for the 6-$j$ 
symbols. The corresponding calculation in the covariant approach involves 
considerably more labour. Also, the result ``$4+14=9+9$'' for doublet to 
doublet one-pion transitions is a general result which again can easily 
be derived using the 6-$j$ approach\cite{santafe1}.

\newpage

\section{Constituent Quark Model Approach to Light-Side Transitions}

Interest in the constituent quark model has recently been rekindled by the 
discovery (or rediscovery) that two-body spin-spin interactions between 
quarks are non-leading in $1/N_C$, at least in the baryon 
sector~\cite{santafe4}. Thus, to leading order in $1/N_C$, light quarks 
behave as if they were heavy as concerns their spin interactions.
In the constituent quark model approach one further
assumes that spin and orbital degrees of freedom decouple. One can
therefore classify the light diquark system 
in terms of $SU(2N_f)\otimes O(3)$ symmetry multiplets. 
Transitions between light quark systems are parametrized in terms of a set 
of one-body operators whose matrix elements are then evaluated between the 
$SU(2N_f)\otimes O(3)$ multiplets.

The constituent quark model light-side spin wave functions are
constructed according to the coupling scheme
\begin{equation}\label{eqn10}
{\textstyle\frac12^+}\otimes{\textstyle\frac12^+}\otimes l_K\otimes l_k
\Rightarrow j^P
\end{equation}
where $l_K$ and $l_k$ denote the two possible orbital angular momenta of
the light diquark with parities $(-1)^{l_K}$ and $(-1)^{l_k}$, respectively.
The construction of the light-side spin wave functions proceeds in complete
analogy to the atomic helium problem only that one has to take into
account the extra colour and flavour degrees of freedom present in the
quark case.
Table 1 lists the appropiate light-side spin wave functions in covariant
form. Again, the corresponding C.G. coupling expressions can easily be
obtained from the covariant expressions by the appropiate $m$-quantum number
projections.

Let us illustrate how the constituent quark model for the light-side
transitions leads to predictions that go beyond the HQS predictions
by calculating the ground state to ground state current transitions.
The relevant light-side one-body transition operator is given by
\begin{equation}\label{eqn11}
O=A(\omega)\cdot\one\otimes\one
\end{equation}
which has to be evaluated between the ground state diquark spin wave functions.
There are altogether three ground state to ground state heavy baryon
form factors or 
Isgur-Wise functions, one for the $\Lambda_b\rightarrow\Lambda_c$ 
transition and two for the $\{\Sigma_b\}\rightarrow\{\Sigma_c\}$ 
transitions. Equation~(\ref{eqn11}) tells us that they can all be expressed 
in terms of the single form factor $A(\omega)$, where $A(1)=1$ at zero 
recoil. One then finds that the current transition amplitudes are given 
by~\cite{santafe1,santafe5,santafe6}
\begin{eqnarray}
\Lambda_b\rightarrow\Lambda_c
  &:&M^\lambda=\bar u_2\Gamma^\lambda u_1\frac{\omega+1}2A(\omega)\\
  \{\Sigma_b\}\rightarrow\{\Sigma_c\}
  &:&M^\lambda=\bar\psi_2^\nu\Gamma^\lambda\psi_1^\mu
  (-\frac{\omega+1}2g_{\mu\nu}+\frac12v_1^\nu v_2^\mu)A(\omega)\nonumber
\end{eqnarray}
The same result has been obtained by C.K.Chow by analyzing the large
$N_C$ limit of QCD~\cite{santafe7}.

For the current transitions from the bottom baryon ground states to the
$p$-wave charm baryon states one similarly 
reduces the number of reduced form factors when invoking
$SU(2N_f)\otimes O(3)$ symmetry in addition to HQS. For the transition into 
the $K$-multiplet one has a reduction from five HQS reduced form factors 
to two constituent quark model form factors  
whereas for transitions into the $k$-multiplet one can relate two HQS 
reduced form factors to one single spin-orbit form
factor~\cite{santafe5}. These are
testable predictions in as much as the population of helicity states in
the daughter baryon is fixed resulting in a characteristic decay pattern
of its subsequent decay.

The one-pion and photon transitions can be 
treated in a similar manner. Again one finds a significant simplification 
of the HQS structure, i.e. the number of coupling factors is reduced from 
those listed in Eqs.~(\ref{eqn6}) and~(\ref{eqn7}) when 
$SU(2N_f)\otimes O(3)$ is invoked in addition to HQS. Results for the
one-pion transitions can be found in~\cite{santafe8}. Corresponding results
for the photon transitions are presently being worked out. We mention that
the constituent quark model approach leads to a one-pion width of
$\Gamma\cong 1$~MeV for the recently observed charm baryon state 
${\Xi_c^0}^*(2643)$. This width is consistent with the experimental upper 
width limit of $5.5$~MeV but unfortunately is too small to be measured with
present techniques~\cite{santafe9}.
 
\section{Asymmetry Parameters in $\Lambda_c \rightarrow \Lambda_s$
Transitions}

Recently the ARGUS and CLEO collaborations have determined
the asymmetry parameters in the semileptonic transition
$\Lambda_c\rightarrow \Lambda_s e^+ \nu_e$
~\cite{santafe10} and in the nonleptonic
one-pion transition $\Lambda_c \rightarrow \Lambda_s + \pi^+$
~\cite{santafe11}.
In both cases the measured asymmetry parameter $\alpha$
(or, equivalently, the polarization of the daughter 
baryon $\Lambda_s$) turns out to be rather close to
$-1$. The question is whether these two results have a common theoretical
explanation. Since the literature contains some wrong statements
on this issue I
want to take the opportunity to clarify the situation.

To begin with let me remind you that there is a remarkable
prediction of HQS for the heavy to light semileptonic transition
$\Lambda_c \rightarrow \Lambda_s$ at zero momentum transfer $q^2=0$. The
$\Lambda_s$ is predicted to emerge with 100\% negative
polarization at this point~\cite{santafe12}. Within error bars this is borne out
by experiment~\cite{santafe10}. All what is needed in this prediction of HQS is a heavy
$\Lambda_c$ while the $\Lambda_s$ can be taken to be light.

Let me assume for the moment  that the nonleptonic decay
$\Lambda_c \rightarrow \Lambda_s + \pi^+$
is dominated by the
so-called factorizable contribution (diagrams I in Fig.3). If this were
the case the
asymmetries in the two decays would in fact become related. Let me,
however, hasten to add beforehand
that the nonleptonic charm baryon decays are not dominated by the
factorizable diagram as we shall presently see. Returning to the
factorizable contribution in Fig.3 one might wonder why there would
be a relation at all between two different components of the weak
$c \rightarrow s$
current: the nonleptonic one-pion decay tests the scalar current
component whereas
in the semileptonic transition one is testing the longitudinal current 
component. A priori these two components are not related except at the
point $q^2=0$. This can be seen by projecting the relevant
current components using the appropiate polarization four-vectors. For 
these one has
\begin{eqnarray}
\mbox{longitudinal:}\qquad\epsilon_\mu(0)
  &=&\frac1{\sqrt{q^2}}(|\vec q|,0,0,q_0)\nonumber\\
\mbox{scalar:}\qquad\epsilon_\mu(s)
  &=&\frac1{\sqrt{q^2}}(q_0,0,0,|\vec q|)\label{eqn12}
\end{eqnarray}
where one should keep in mind that the transverse pieces of the vector
current transitions decouple at $q^2=0$. From Eq.~(\ref{eqn12}) it is evident 
that $\epsilon_\mu(0)=\epsilon_\mu(s)$ at $q^2=0$ where $|\vec q|=q_0$
and thus the scalar and 
longitudinal components become related at this point. The pion point 
$q^2=m_\pi^2$ is so close to $q^2=0$ that the extrapolation to $q^2=0$ is 
perfectly save. With what has been said up to now it is then very tempting to 
(erraneously!) invoke a common theoretical Heavy Quark Symmetry origin for 
the near equality of the above two asymmetries.

As concerns the mesonic sector one knows
that the one-pion transitions and semileptonic transitions close
to $q^2=0$ are in fact related.
However, nonleptonic baryon
decays are quite different from nonleptonic mesonic decays
in that there are more contributing diagrams in the baryon case.
In addition to the factorizing diagrams I in Fig.3 there are
the nonfactorizing diagrams II$_{a,b}$ and III in Fig.3. That the
nonfactorizing diagrams
cannot be neglected in charm baryon decays can be surmized from the fact
that some of the
observed nonleptonic charm baryon decays can only proceed via the
nonfactorizing diagrams. As a sample decay take the decay 
$\Lambda_c \rightarrow \Xi^0 + K^+$ which proceeds through diagrams II$_a$ and III
and yet has a
sizeable experimental branching fraction. From all what has been
said one must conclude 
that the observation of a near maximal negative polarization in the decay
$\Lambda_c \rightarrow \Lambda_s + \pi$ does not have a simple explanation
but must be considered to be a dynamic accident resulting from the
interplay of a number of contributing diagrams.

\section{Some Selected Remarks on Exclusive Nonleptonic Bottom Baryon Decays}

At the Br\"ussel '95 EPS meeting the ALEPH~\cite{santafe13} and
DELPHI~\cite{santafe14} collaborations presented preliminary evidence for the 
nonleptonic decay $\Lambda_b \rightarrow \Lambda_c + \pi^-$. Projecting into 
the future one can imagine that, given enough statistics, the full decay chain
\begin{equation}
\Lambda_b\rightarrow\hbox{\vtop{%
  \hbox{$\Lambda_c+\pi^-$}
  \hbox{$\ \hookrightarrow\hbox{\vtop{%
    \hbox{$\Lambda_s+\pi^+$}
    \hbox{$\ \hookrightarrow p+\pi^-$}}}$}}}
\label{eqn13}
\end{equation}
can eventually be reconstructed. The angular decay distribution
in the decay chain can be seen to be given by~\cite{santafe15}
\begin{equation}\label{eqn14}
W(\theta_2,\theta_3)=1+\alpha_1\alpha_2\cos\theta_2
  +\alpha_3(\alpha_2+\alpha_1\cos\theta_2)\cos\theta_3
\end{equation}
where $\alpha_1$, $\alpha_2$ and $\alpha_3$ are the asymmetry parameters
in the decays
$\Lambda_b \rightarrow \Lambda_c + \pi^-$,
$\Lambda_c \rightarrow \Lambda_s + \pi^+$ and
$\Lambda_s \rightarrow p +\pi^-$, respectively.
The polar angles $\theta_2$ and $\theta_3$ are defined through the
momenta of the $\Lambda_s$ in the $\Lambda_c$ rest frame, and the proton
in the $\Lambda_s$ rest frame,respectively. If one integrates over 
$\cos\theta_3$ one arrives at the the angular decay distribution
\begin{equation}\label{eqn15}
W(\theta_2)=1+\alpha_1\alpha_2\cos\theta_2
\end{equation}
Since $\alpha_2$ has been measured ($\alpha_2=-0.89{+0.19\atop-0.12}$) the 
decay distribution Eq.(\ref{eqn15}) can be used to  determine the unknown 
asymmetry parameter $\alpha_1$ in the decay 
$\Lambda_b \rightarrow \Lambda_c + \pi^-$.
The quality of this measurement
depends on the accuracy with which the $c\rightarrow s$ asymmetry parameter
$\alpha_2$ is known. This simple observation invites a general comment:
the quality of the analysis of future $b\rightarrow c$ data depends on the
quality of the present $c\rightarrow s$ physics analysis. This holds both for
polarization type variables and also for branching
ratios. The message for experimentalists
is obvious: try to improve on the error bars in charm decays if only
for the sake of improving future bottom decay analysis. One step further down
in the decay chain the situation is quite satisfactory in this regard.
The asymmetry parameter $\alpha_3$ in the decay
$\Lambda_s \rightarrow p +\pi^-$ (which is used
as an analyzer to determine the asymmetry parameter $\alpha_2$
in the decay $\Lambda_c \rightarrow \Lambda_s + \pi^+$) is known with
sufficient accuracy ($\alpha_3=0.642\pm 0.013$) not to limit the accuracy 
of the $\alpha_2$ determination.

Where do we stand at the moment in our understanding of exclusive nonleptonic
bottom baryon decays such as the decay
$\Lambda_b \rightarrow \Lambda_c + \pi^-$? As a 
theoretician I must ruefully admit that there exist no satisfactory
theory for exclusive nonleptonic bottom baryon decays at present.
This is in marked difference to exclusive nonleptonic
bottom meson decays where one has achieved a basic understanding
over the last few years. The reason for this was stated before:
bottom meson decays can be described by calculable factorizing    
contributions whereas heavy baryon decays involve also nonfactorizing
contributions which are difficult to calculate. 

One may turn to the strange and charm baryon sector for advice.
Nonleptonic hyperon decays have been traditionally calculated using
the three ingredients " soft pion theorem + current algebra + nearest
pole dominance ". This approach is not easily carried over to the
$c \rightarrow s$ and $b \rightarrow c$ sectors in as much as the
pion is no longer really soft in the latter two cases. For example,
in the decay $\Lambda_c \rightarrow \Lambda_s + \pi^+$ one has
$|\vec p_\pi|=0.86$~GeV and in the decay
$\Lambda_b \rightarrow \Lambda_c + \pi^-$ one has $|\vec p_\pi|=2.21$~GeV.
The soft-pion approach may barely
be justified in  $c \rightarrow s$ decays but certainly does not make
sense for the $b \rightarrow c$ decays. A related problem is that
the energy released in the decays is so large that one is far away
from the region where the ground state baryons can be used for the
pole dominance approximation. Nevertheless this approach has been
applied with reasonable success to the $c \rightarrow s$ decays but
certainly should not be used for bottom baryon decays. 

One can then ask oneself whether there is any reason to believe that 
in bottom baryon decays the nonfactorizing contributions are suppressed.
In such a case one could then hope to have a theoretical handle on nonleptonic
bottom baryon decays. Turning to $1/N_C$ arguments does not help.
Although the nonfactorizing diagrams II$_{a,b}$ and III appear to be
colour suppressed relative to the factorizing diagram I$_a$ this is true
only for $N_C=3$. Considering the fact that baryons contain $N_C$
quarks as $N_C\rightarrow\infty$ with $O(N_C)$ light flavoured
quarks there is a combinatorial factor proportional to $N_C$ which
cancels the explicit diagrammatic $1/N_C$ factor. This result is in
agreement with the analysis in the charm and strangeness sector where the
nonfactorizing contributions are certainly needed.
However, there do in fact exist qualitative arguments for a suppression
of the nonfactorizing diagrams in nonleptonic bottom baryon decays.
First of all diagram
II$_b$ can be seen to be helicity suppressed~\cite{santafe16}.
Second, in diagrams II$_a$ and III
one needs to create an energetic light quark-antiquark pair from the vacuum.
Since there is a considerable amount of energy released e.g. in the decay
$\Lambda_b \rightarrow \Lambda_c + \pi^-$  both suppression
mechanisms should be quite effective.
For example, from the remaining factorizing contribution one would then 
predict that the asymmetry parameter $\alpha_1$ in the decay
$\Lambda_b \rightarrow \Lambda_c + \pi^-$
is maximally negative following the HQS arguments presented in Sec.6.
Needless to say that it would be highly desirable to put these
qualitative arguments 
on a more quantitative basis.

\section{Concluding Remarks}

In this review we have limited our attention to the exclusive decay
modes of heavy baryons. This choice was dictated by space limitations.
There have certainly been some important theoretical advances in the 
understanding of the inclusive decays of heavy baryons which I did not
have time to cover. These advances are important since they have a bearing
on two unresolved puzzles in bottom baryon physics that have been widely
publicized in the last year. The first puzzle concerns the unexpectedly
small polarization of the $\Lambda_b$ from Z-decays as measured by the
ALEPH collaboration~\cite{santafe17}. The second puzzle concerns
the sizeable deviation of
the lifetime of the $\Lambda_b$ from the life time average of the
other bottom hadrons. The latter topic was covered by Bijan Nemati at this 
workshop.
\vspace{1cm}\\
{\bf Acknowledgement:}\\
Much of the material presented in this review is drawn from work done in 
collaboration with F.~Hussain, M.~Kr\"amer, J.~Landgraf, D.~Pirjol and 
S.~Tawfiq. I would like to express my gratitude to S.~Groote
for help and advice on the LATEX version of this report.

\vspace{1cm}
\centerline{\bf\Large Figure Captions }
\vspace{0.5cm}
\newcounter{fig}
\begin{list}{\bf\rm Fig.\ \arabic{fig}: }{\usecounter{fig}
\labelwidth1.6cm \leftmargin2.5cm \labelsep0.4cm \itemsep0ex plus0.2ex }

\item Generic picture of bottom to charm current transitions, and 
  pion and photon transitions in the charm sector in the HQS limit 
  $m_Q\rightarrow\infty$
\item One-pion transition strengths for the transitions 
  $\{\Lambda_{QK2}^{**}\}\rightarrow\{\Sigma_Q\}+\pi$. 
  Degeneracy levels are split for illustrative purposes.
\item Quark flow diagrams for exclusive nonleptonic baryon decays. The 
  explicit quark flavour labelling is for the decay 
  $\Lambda_c\rightarrow\Lambda_s+\pi^+$.

\end{list}

\newpage

\begin{table*}
\begin{center}
\begin{tabular}{|c|cccc|}
\hline
& \begin{tabular}{c}light side s.w.f.\\${\hat\phi}^{\mu_1\cdots\mu_j}$
  \end{tabular}
& $j^P$
& \begin{tabular}{c}heavy side s.w.f.\\$\psi_{\mu_1\cdots\mu_j}$
  \end{tabular}
& $J^P$
\\ \hline\hline
\multicolumn{5}{|l|}{$s$-wave states ($l_k=0$, $l_K=0$)}\\
  $\Lambda_Q$&$\hat\chi$&$0^+$&$u$&$\frac12^+$
\\ \hline
  $\{\Sigma_Q\}$
  & ${\hat\chi}^{1\mu_1}$&$1^+$
  & $\begin{array}{c}\frac1{\sqrt3}\gamma^\perp_{\mu_1}\gamma_5u\\
      u_{\mu_1}\end{array}$
  & $\begin{array}{c}\frac12^+\\\frac32^+\end{array}$
\\ \hline\hline
\multicolumn{5}{|l|}{$p$-wave states ($l_k=0$, $l_K=1$)}\\
  $\{\Lambda_{QK1}^{**}\}$
  & ${\hat\chi}^0K_\perp^{\mu_1}$&$1^-$
  & $\begin{array}{c}\frac1{\sqrt3}\gamma^\perp_{\mu_1}\gamma_5u\\
      u_{\mu_1}\end{array}$
  & $\begin{array}{c}\frac12^-\\\frac32^-\end{array}$ 
\\ \hline
  $\Sigma_{QK0}^{**}$
  & $\frac1{\sqrt3}{\hat\chi}^1\cdot K_\perp$&$0^-$&$u$&$\frac12^-$
\\ \hline
  $\{\Sigma_{QK1}^{**}\}$
  & $\frac i{\sqrt2}\varepsilon(\mu_1{\hat\chi}^1K_\perp v)$&$1^{-}$
  & $\begin{array}{c}\frac1{\sqrt3}\gamma^\perp_{\mu_1}\gamma_5u\\
      u_{\mu_1}\end{array}$
  & $\begin{array}{c}\frac12^-\\\frac32^-\end{array}$
\\ \hline
  $\{\Sigma_{QK2}^{**}\}$&
  $\frac12\{{\hat\chi}^{1,\mu_1}K_\perp^{\mu_2}\}_{0}$&$2^-$
  & $\begin{array}{c}\frac1{\sqrt{10}}\gamma_5\gamma^\perp_{\{\mu_1}
      u_{\mu_2\}_0}^{\phantom{\perp}}\\u_{\mu_1\mu_2}\end{array}$
  & $\begin{array}{c}\frac32^-\\\frac52^-\end{array}$
\\ \hline\hline
\multicolumn{5}{|l|}{$p$-wave states ($l_k=1$, $l_K=0$)}\\
  $\{\Sigma_{Qk1}^{**}\}$&${\hat\chi}^0k^{\mu_1}_\perp$&$1^-$
  & $\begin{array}{c}\frac1{\sqrt3}\gamma^\perp_{\mu_1}\gamma_5u\\
      u_{\mu_{1}}\end{array}$
  & $\begin{array}{c}\frac12^-\\\frac32^-\end{array}$
\\ \hline
  $\Lambda_{Qk0}^{**}$&
  $\frac1{\sqrt3}{\hat\chi}^1\cdot k_\perp$&$0^-$&$u$&$\frac12^-$
\\ \hline
  $\{\Lambda_{Qk1}^{**}\}$
  & $\frac i{\sqrt2}\varepsilon(\mu_1{\hat\chi}^1k_\perp v)$&$1^-$
  & $\begin{array}{c}\frac1{\sqrt3}\gamma^\perp_{\mu_1}\gamma_5u\\
      u_{\mu_{1}}\end{array}$
  & $\begin{array}{c}\frac12^-\\\frac32^-\end{array}$
\\ \hline
  $\{\Lambda_{Qk2}^{**}\}$
  & $\frac12\{{\hat\chi}^{1,\mu_1}k_\perp^{\mu_2}\}_0$&$2^-$
  & $\begin{array}{c}\frac1{\sqrt{10}}\gamma_5\gamma^\perp_{\{\mu_1}
      u_{\mu_2\}_0}^{\phantom{\perp}}\\u_{\mu_1\mu_2}\end{array}$
  & $\begin{array}{c}\frac32^-\\\frac52^-\end{array}$
\\ \hline
\end{tabular}
\end{center}
\caption{Spin wave functions (s.w.f.) of heavy $\Lambda$-type 
  and $\Sigma$-type $s$- and $p$-wave heavy baryons 
  ($\hat\chi^0=\frac1{2\sqrt 2}[(\slv+1)\gamma_5C]$;
  $\hat\chi^1_\mu=\frac1{2\sqrt 2}[(\slv+1)\gamma^\perp_\mu C]$).}
\vspace{5mm}
\end{table*}
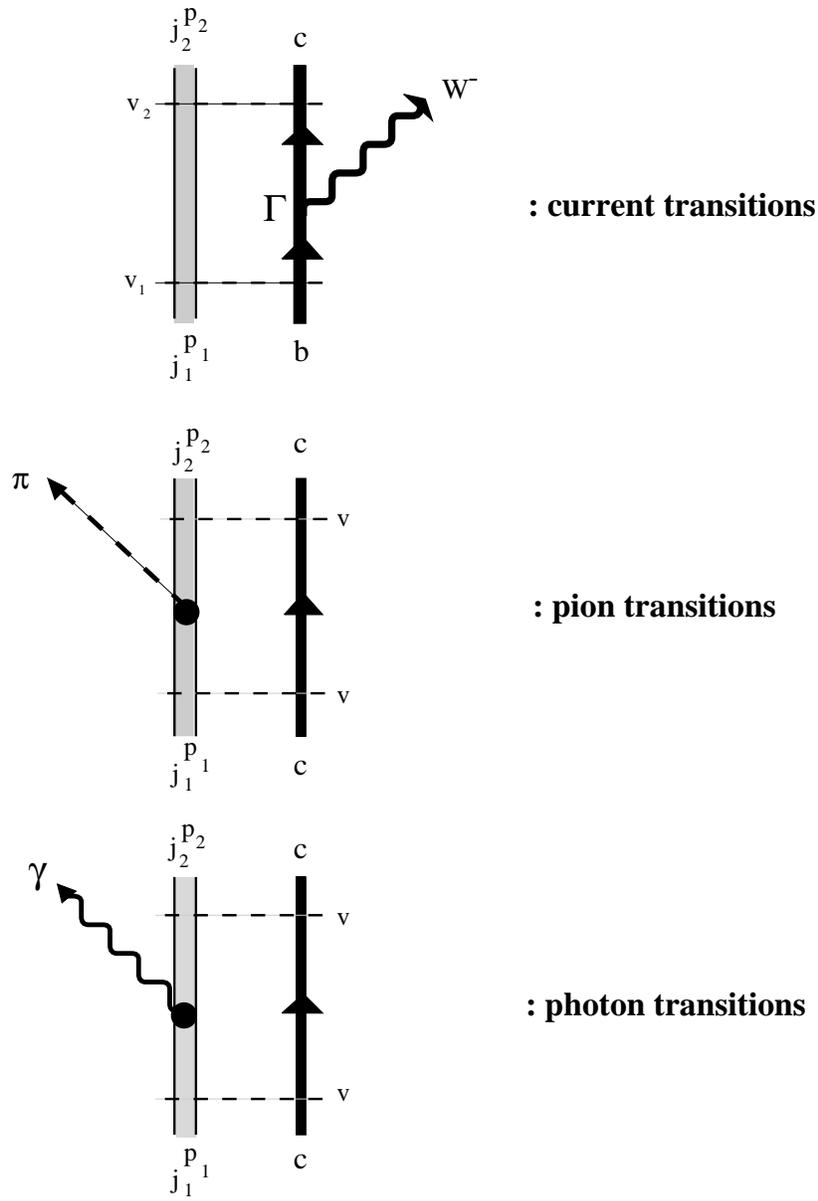
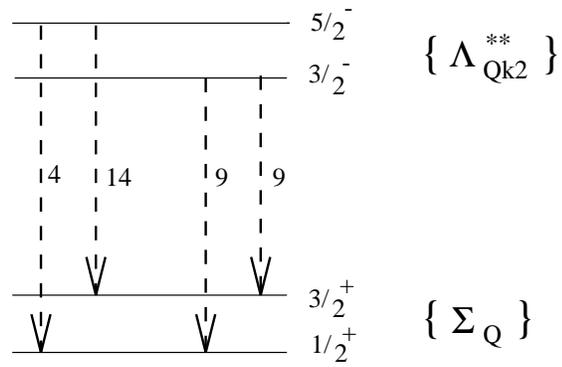
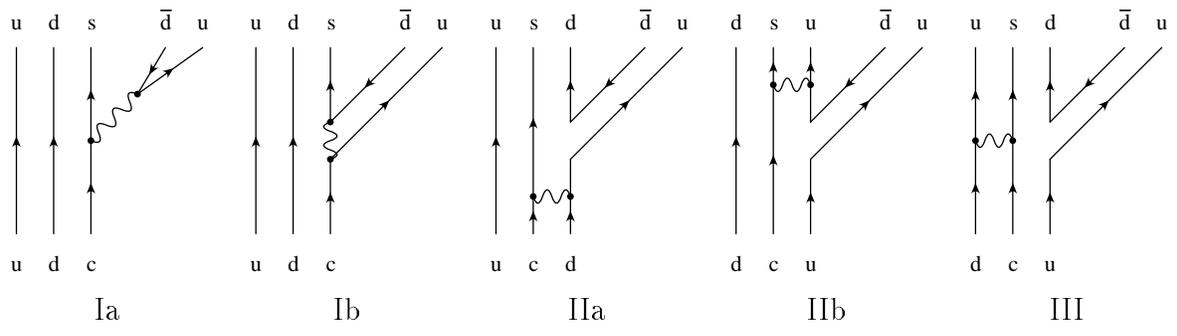
\end{document}